\documentclass[%
 reprint,
 amsmath,amssymb,
 aps,
floatfix,
]{revtex4-1}
\usepackage[T1]{fontenc}
\usepackage{bbold}
\usepackage{graphicx}
\usepackage{dcolumn}
\usepackage{bm}
\usepackage{color}
\usepackage{transparent}
\usepackage{hyperref}

\newcommand{\mbf}[1]{\mathbf{#1}}

\newcommand{\ket}[1]{\left| #1 \right\rangle}
\newcommand{\bra}[1]{\left\langle #1 \right|}

\newcommand{\Tr}{\mathrm{Tr}}

\definecolor{orange}{rgb}{1,0.66,0}

\begin{document}


\title{Quantum computational supremacy in the sampling of bosonic random walkers on a one-dimensional lattice}

\author{Gopikrishnan Muraleedharan}
  \email{gopu90@unm.edu}
\author{Akimasa Miyake}
\author{Ivan H. Deutsch}%

\affiliation{%
Center for Quantum Information and Control (CQuIC), Department of Physics and Astronomy , University of New Mexico
}%

\date{\today}

\begin{abstract}
We study the sampling complexity of a probability distribution associated with an ensemble of identical noninteracting bosons undergoing a quantum random walk on a one-dimensional lattice.   With uniform nearest-neighbor hopping we show that one can efficiently sample the distribution for times logarithmic in the size of the system, while for longer times there is no known efficient sampling algorithm. With time-dependent hopping and optimal control, we design the time evolution to approximate an arbitrary Haar-random unitary map analogous to that designed for photons in a linear optical network.  This approach highlights a route to generating quantum complexity by optimal control only of a single-body unitary matrix. We study this in the context of two potential experimental realizations: a spinor optical lattice of ultracold atoms and a quantum gas microscope.  

\end{abstract}

\maketitle
\section{Introduction}
What level of quantum complexity cannot be efficiently simulated on a classical computer?  This is a central question in quantum information science that impacts fundamental physics from the nature of condensed matter~\cite{Amico2008,Cirac2012,condensedmatter} to the geometry of space time~\cite{gravity}. While it is widely believed that a universal fault-tolerant quantum computer achieves such complexity, implementation of such a device is still a distant prospect. Nonetheless, more modest devices designed for  a limited task could supersede the power of a classical computer and achieve so-called ``quantum computational  supremacy''~\cite{supremacy,Harrow2017} (QCS). In particular, a quantum device can yield random outcomes sampled from a probability distribution such that no classical computer could efficiently simulate its statistics. In their seminal paper, Aaronson and Arkhipov~\cite{linoptics} showed that QCS could arise from ``sampling complexity''  in the most unlikely of places: linear optics. Based on the hierarchy of computational complexity classes, we believe that the number distribution of identical photons scattering from a linear optical network cannot be efficiently simulated. In this case, the nonclassical behavior arises solely from the quantum statistics of identical particles, or equivalently, the entanglement between modes, since photons are noninteracting particles in linear optics. The study of Boson Sampling that achieves quantum supremacy is a goal pursued worldwide~\cite{Spring2013,gogolin2013,Crespi2013,Tillmann2013,Broome2013,Lund2014,Tichy2014,Rahimi-Keshari2015,Tillmann2015,Gard2015,Olson2015,Rahimi-Keshari2016,Drummond2016,Aaronson2016,peropadre2017,Wang2017,He2017,Loredo2017,Neville2017,Barkhofen2017}.

In this work we extend the physical paradigm for studying Boson Sampling to a continuous-time quantum random walk of noninteracting identical bosons on a 1D lattice~\cite{Underwood2012,Lahini2012,Sansoni2012,Childs2013,Qin2014,Preiss2015}. This problem was studied recently by Deshpande {\em et. al}~\cite{Deshpande2017} who demonstrated how one might consider sampling complexity as a new type  of order parameter in a dynamical phase transition. In particular, using Lieb-Robinson bounds, they showed how sampling transitions from ``easy" to ``hard " as a function of time in the evolution. 

Following these works, our motivation is here two-fold. Firstly, we seek to understand the minimal complexity necessary to achieve QCS. A 1D gas of noninteracting bosons is perhaps the simplest nontrivial many-body quantum system.  Secondly, this paradigm can be realized in physical platforms that might lead to more scalable implementations, e.g., ultracold bosonic atoms in an optical lattice. Recent experiments with spinor lattices ~\cite{oplatt3,Karski2011,Mandel2003,Mischuck2010}, optical tweezers~\cite{kaufman2012,Kaufman2014,Norcia2018,Cooper2018}, and quantum gas microscopes~\cite{atomicgas,Sherson2010} demonstrate the possibility of observing controlled transport of modest sized ensembles of identical atoms and direct readout of the atom number distribution at the lattice sites~\cite{atomicgas,Sherson2010,Kaufman2014,Kaufman2016,Lester2017,Kaufman2018}. In contrast to linear optics, imperfections such as photon loss, misalignment of a large interferometer, detection efficiency, and the challenge of producing a large input $(\sim 50)$ of identical bosons might be mitigated, improving the prospect of near-term demonstration of QCS. The ability to prepare, address, and measure individual atoms is a powerful toolbox that has enabled explorations in quantum transport, including the Hong-Ou-Mandel effect~\cite{Kaufman2014} and more general many-body interference~\cite{Kaufman2018}, the relationship of entanglement entropy and thermalization entropy~\cite{Islam2015, Kaufman2016}, and many-body localization~\cite{Schreiber2015}. This capability makes cold atoms in optical traps a natural platform for exploring sampling complexity as a path to QCS.

The remainder of this article is organized as follows.  In Sec. II we define the problem of Boson Sampling for identical bosonic quantum random walkers on a lattice. In particular, we consider first the case of a 1D static lattice with uniform nearest-neighbor hopping, well known in physics as the tight-binding model for transport in a periodic lattice.  In Sec. III we continue to restrict to nearest neighbor hopping, but we generalize to allow for a time-dependent lattice. We show how for different geometries, the system can be made ``controllable," and using quantum optimal control, we show how one can engineering waveforms to implement the complexity that is conjectured to be necessary to ensure QCS.  We summarize and give an outlook to future research in Sec. IV.

\section{Boson Sampling in the static 1D tight-binding model}
In Boson Sampling one considers $N$ identical noninteracting bosons in $M$ modes evolving by the linear map, $U a^{\dagger}_{l'} U^{\dagger} = \sum\limits_{l} \Lambda_{l' l} a^{\dagger}_{l}$
where $\Lambda$ is an $M\times M$ unitary matrix, which we denote as the ``transition matrix."  Given an input Fock state $\ket{\mbf{n}^{in}} = \ket{n^{in}_1,n^{in}_2,\dots,n^{in}_M}$, the transition probability to a corresponding output Fock state with $\sum_l n^{out}_l = \sum_l n^{in}_l=N$ is
\begin{equation}
 P(\mbf{n}^{out}|\mbf{n}^{in}) =|\bra{\mbf{n}^{out}} U \ket{\mbf{n}^{in}}|^2 =  \frac{|\text{Perm}(\Lambda_{\mbf{n}^{out}|\mbf{n}^{in}})|^2}{n^{in}_1! .. n^{in}_M!\; n^{out}_1! .. n^{out}_M!},
\end{equation}
Where $\Lambda_{\mbf{n}^{out}|\mbf{n}^{in}}$ is a ``submatrix'' obtained by repeating ${n}^{in}_{l}$ times the $l^{th}$ column of $\Lambda$, and ${n}^{out}_{l'}$ times its $l'^{th}$ row. $\text{Perm}(\Lambda)$ is the permanent of the matrix.  We restrict our attention here to a fiducial input state with 1 boson in each of $N<M$ input modes and 0 bosons in the remaining  $M-N$ modes. A Boson Sampler then outputs a random sequence $\mbf{n}^{out}$ with probability $P(\mbf{n}^{out})$.  The complexity of sampling from this distribution on a classical computer depends on the complexity of calculating a multiplicative approximation to the permanent of $\Lambda$ and its submatrices~\cite{VALIANT1979}.  

In the optical Boson Sampling setting, $U$ is the S-matrix for scattering incoming photons into the outgoing modes of a linear optical network~\cite{linoptics}.  There, an arbitrary unitary matrix $\Lambda_{l'l}$ can be constructed through a sequence of phase shifters and beam splitters~\cite{zeilinger}. Here we consider a continuous time quantum random walk of identical bosons on a lattice, described by the second quantized Hamiltonian $H=\sum_{l',l} h_{l'l} a^\dag_{l'} a_l$,
where $h_{l'l}$ is the generator of tunneling (hopping) from one lattice site $l$ to another $l'$. The unitary evolution of creation operators is exactly the same as for linear optics, except that the S-matrix is now replaced by the time evolution operator, 
$U(t) =\sum_{l',l}  u_{l'l}(t) a^\dag_{l'} a_l$, where $u_{l'l}(t) = \left(e^{-i h t}\right)_{l'l}$ is the single particle time evolution operator in the lattice site basis. Setting $\Lambda_{l'l}=u_{l'l}(T)$ for the final time $T$, we see the isomorphism between the quantum random walk and the linear optical network.  

If the Hamiltonian $h_{l'l}$ describes nonlocal hopping on an arbitrary graph, then one can obtain an arbitrary $\Lambda$. Here we restrict our attention to  nearest-neighbor hopping.  In the simplest case of uniform hopping $h_{l'l} = -J(\delta_{l
',l+1}+\delta_{l',l-1})$; we consider periodic boundary conditions. The Hamiltonian is trivially diagonalized in the Bloch basis with creation operators $b^{\dagger}_q = \frac{1}{M} \sum\limits_{l} e^{-i l \frac{2\pi q}{M}} a^{\dagger}_{l}$, yielding the band structure in the tight-binding model and resulting time evolution operator
\begin{equation}
U(t) = \sum\limits_q e^{i Jt \cos(\frac{2\pi q}{M})} b^\dagger_q b_{q} .
\end{equation}
While the time evolution is trivial, the same cannot be said in general for sampling complexity. In particular, sampling depends on the basis in which we measure, because the permanent is {\em basis dependent}.  Transforming back to the lattice site basis, we have $U(t) =\sum\limits_{l,l'} u_{ll'}(t) a^\dag_l a_{l'}$ where
\begin{equation}
u_{ll'}(t) =\frac{1}{M^2} \sum_q e^{-i(l-l') \frac{2\pi q}{M}} e^{i Jt \cos(\frac{2\pi q}{M})}.
\end{equation}
The protocol for Boson Sampling in this system is as follows.  A single boson is prepared at each of $N$ sites of an $M$-site periodic lattice, each in an identical vibrational level.  After some final time $T$ we measure the number of bosons in each of the sites, which provides a sample from a probability distribution that depends on the permanent of $\Lambda_{ll'}=u_{ll'}(T)$. We seek to understand the computational complexity of calculating this permanent.

\begin{figure}
\centering
\includegraphics[width=\columnwidth]{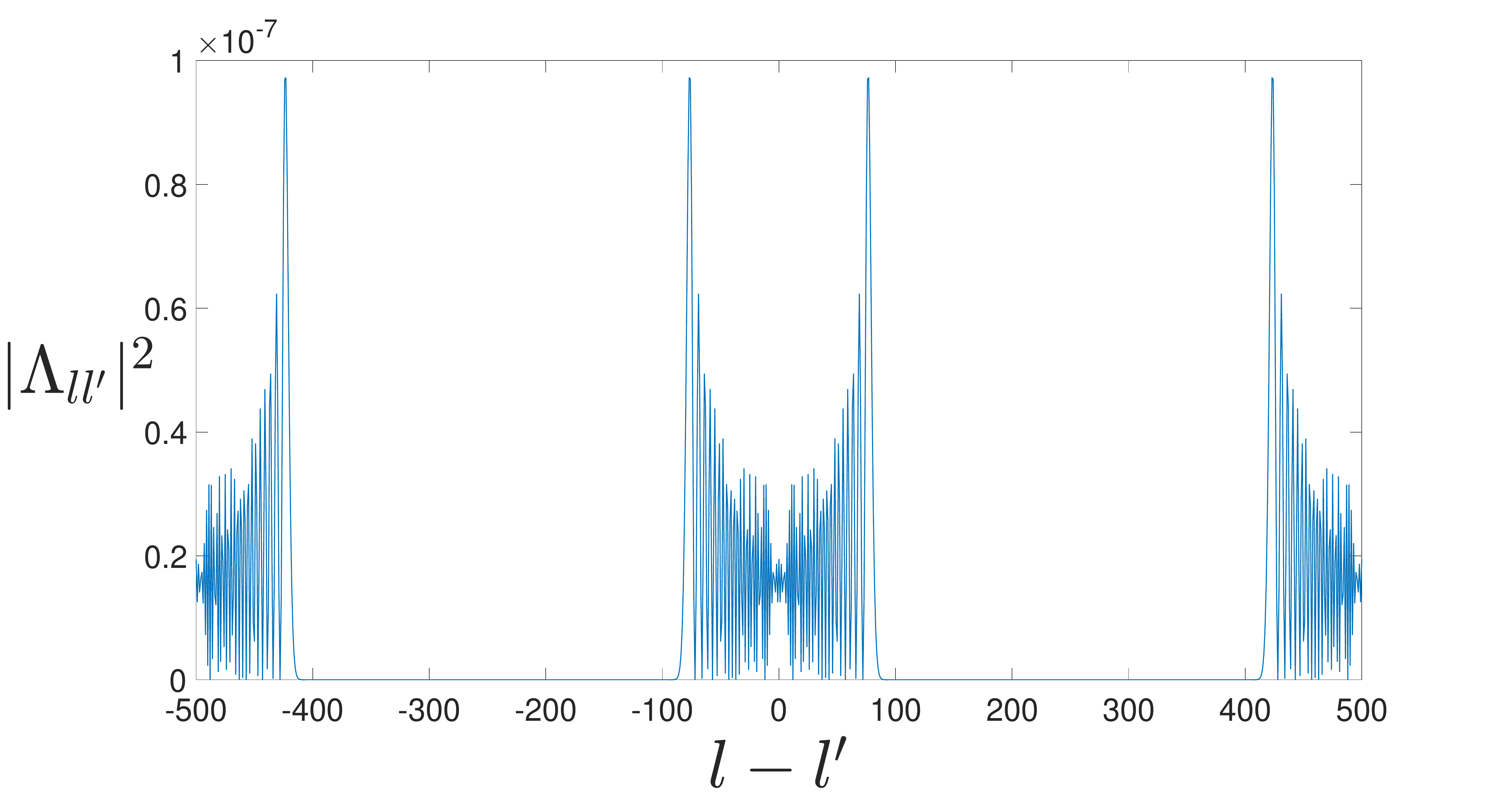}
\caption{Absolute squared values of the transition matrix elements, $\vert \Lambda_{ll'}\vert^2$, as function of $l-l'$, for $M$=500 lattice sites with periodic boundary conditions, after a  time $T = 80/J$. The transition probability reflects the ballistic tunneling of a quantum random walker. }
\label{fig:matrixplot}
\end{figure}

The matrix $\Lambda_{ll'}$ has a structure imposed by the restricted nature of the Hamiltonian. It is a circulant matrix whose elements depend only on $(l-l') \ \text{mod} \  M$. In a translationally invariant system, the transition amplitude depends only on the ballistic distance traveled. A plot of the transition amplitudes in Fig.~\ref{fig:matrixplot} shows the familiar wave function for a continuous-time quantum random walk ~\cite{Qwalk1}.  Note that the elements of the matrix decays exponentially after a ``band" of elements, and again increase towards edges due to the periodic boundary conditions. We call this band $B$, which is proportional to $T$.  We thus consider a transition matrix $\Lambda'$  and set $\Lambda'_{ll'}=0$ for $|\Lambda_{ll'}|<\epsilon \vert| \Lambda\vert|$ for some $\epsilon \ll 1$, where  $\vert| \Lambda\vert|$ is a norm of the matrix. The transition matrix $\Lambda'$ then takes the form of a doubly banded matrix.

One criterion for approximate Boson Sampling given by Aaronson and Arkhipov is that the variation distance between two probability distributions $P(X)$ and $Q(X)$, defined as 
\begin{equation}
    \| P-Q \|=\frac{1}{2} \sum_X |P(X)-Q(X)|,
\end{equation}
should be small.  Here the summation runs over every event.  This condition can be achieved by a multiplicative approximation in probabilities of each of the events.
\begin{eqnarray}
|P(X)-Q(X)| &\leq& \epsilon P(X) \\
 \| P-Q \|&=&\frac{1}{2} \sum_X |P(X)-Q(X)| \nonumber \\ 
 &\leq& \epsilon \sum_X P(X) \leq \epsilon \nonumber
 \end{eqnarray}The total variation distance between the exact occupation number probability distribution of quantum random walkers generated by $\Lambda$ and that generated by $\Lambda'$ will be $O(\epsilon)$. To see this, note that
\begin{eqnarray}
    && \text{Perm}(\Lambda) = \text{Perm}(\Lambda') + \text{Perm}(\Delta) \nonumber\\ &&+  \sum_{k=1}^{n-1}\sum_{s_k, t_k} \text{Perm}(\Delta_{i\in s_k,j\in t_k}) \text{Perm}(\Lambda'_{i\in \bar{s}_k,j\in \bar{t}_k})
\end{eqnarray} 
where $\Delta = \Lambda - \Lambda'$, $s_k$ and $t_k$ are set of $k$ rows or columns respectively, and $\bar{s}$, $\bar{t}$ are their complements~\cite{minc1984}. Keeping only $O(\epsilon)$ terms in the sum,
\begin{eqnarray}
\text{Perm}(\Lambda) &\approx& \text{Perm}(\Lambda') +  \sum_{s_1, t_1} \Delta_{i\in s_1,j\in t_1} \text{Perm}(\Lambda'_{i\in \bar{s}_1,j\in \bar{t}_1})\nonumber\\ 
&\leq& \text{Perm}(\Lambda') + \epsilon \sum_{i,j} \Lambda'_{i,j} \text{Perm}(\Lambda'_{\bar{i},\bar{j}}) \nonumber\\
&=& \text{Perm}(\Lambda') + \epsilon \text{Perm}(\Lambda') \\
|\text{Perm}(\Lambda) &-& \text{Perm}(\Lambda') | \leq \epsilon \text{Perm}(\Lambda')
\end{eqnarray}
Thus, to within multiplicative approximation, we can treat the transition matrix as banded.

In the case where $M$ is sufficiently large compared to $N$, and all of the bosons are trapped in the central region of the lattice, a submatrix of $\Lambda'$ with dimension $N$ is a singly banded Toeplitz matrix. Schwartz~\cite{schwartz} devised an algorithm to calculate the permanent of such matrices which scales as $O({\binom{2B}{B}}^3 \log N)\approx O(2^{6B} \log N)$.  If we include all modes $M$, assuming a lattice on a ring, the transition matrix $\Lambda'$ is a cyclically banded matrix \footnote{A matrix $M$ is cyclically banded if $M_{i,j}=0$, if $N-\omega_2> i-j>\omega_1$ or $N-\omega_1>j-i>\omega_2$, where $N$ is the dimension of matrix $M$. We call $B=\omega_1+\omega_2$, its band}. Cifuentes {\em et. al }~\cite{cifuentes}, showed that one can exactly determine the permanent of such matrices in $O(N 2^{2B})$.  This algorithm does not use the fact that the matrix has a circulant structure, so that the algorithm may be improved to a scaling of $O(\log N 2^{B})$. This follows from the fact that multiplication of $N$ distinct numbers requires $N$ operations whereas calculating $x^N$ only requires $\log(N)$ operations.

Now, if the time of evolution is logarithmic with system size, $T\sim O(\log(N))$, so is $B$. Both the algorithms discussed then scale as $O(\text{poly}(N))$ ~\cite{cifuentes,schwartz}. Thus, up to time of evolution logarithmic in $N$, one can devise an algorithm to efficiently sample from the resulting approximate occupation number probability distribution of quantum random walkers generated by $\Lambda'$. It must be emphasized that we assume no restriction on the initial configuration of bosons, and our bound is also independent of $M$. This is a sharp contrast with an analogous result obtained independently by Deshpande {\em et. al}~\cite{Deshpande2017}. They proved that the variation distance between the exact quantum probability distribution and that of distinguishable particles was exponentially small for times  $t \le 0.9 L/v \sim O(N^{\frac{\beta-1}{d}})$, where $2L$ is the minimum spacing between any two bosons in the initial state \footnote{ The initial configuration of bosons is assumed that they are equally spaced in the lattices. Parametrizing $M=N^{\beta}$, $L=(\frac{M+N}{N})^{\frac{1}{d}}\sim N^{\frac{\beta-1}{d}}$, so $t<O(N^{\frac{\beta-1}{d}})$ is the easy regime.  For example, when $\beta=1$, that is $M\propto N$, our method provides a stronger bound $\log(N)$ instead of constant time.}. This reflects the intuitive fact that for sufficiently short time, the overlap between initially separated bosons is negligible, and thus so is quantum interference.

By contrast, if the time of evolution grows larger than $O(\log N)$, (say $\text{polylog} (N)$ ) the scaling of the algorithms in~\cite{schwartz} and~\cite{cifuentes} becomes super-polynomial (exponential in case of $T\sim O(N))$. To demonstrate QCS it is sufficient that the distribution $P'$ one samples from is a multiplicative approximation to $P$, i.e., $|P'-P|<\epsilon P$.  Unlike for short times, for long times we know of no way to approximate the permanent, nor any multiplicative approximation to the  probability distribution. Other classical sampling algorithms, such as Monte-Carlo sampling, will only allow one to calculate the permanent to within an  {\em additive} approximation. Thus, given the limitations of known methods, for sufficient time of tunneling of quantum random walkers of a 1D lattice, with uniform nearest-neighbor hopping, direct classical simulation of the exact quantum probability distribution based on a multiplicative approximation of the permanent appears to be intractable.

While the known algorithms for calculating the permanent or even approximating it to within a multiplicative error scale poorly, we cannot rule out the possibility of the existence of an efficient classical algorithm for Boson Sampling in this tight-binding periodic lattice.  Nonetheless, given the growth in the graph structure for these matrices~\cite{cifuentes}, we conjecture that this is indeed a hard problem, and noninteracting quantum random walkers show quantum complexity even for this seemingly trivial problem of Bloch band structure in 1D. As further evidence, consider a simulation of the many-body system based on matrix product states (MPS) of bosonic modes~\cite{vidal2003,Temme2012}. We find that the bond dimension of the MPS has a similar scaling to the algorithms described above. For constant time the bond dimension is logarithmic in system size and entanglement growth is polynomial.  For a final time growing linearly with systems size the MPS entanglement grows exponentially. 

\section{Boson Sampling in the 1D time-dependent tight-binding model}

\subsection{Implementing  Haar random unitary transformations via optimal control}
While we conjecture that calculating (within a multiplicative approximation) the permanent of a general banded Topelitz matrix with sufficiently large $B$ is outside the computational complexity class $P$, proving this is beyond the scope of the current work.  A more direct path to QCS in our system is to connect to the arguments of Aaronson and Arkipov~\cite{linoptics}.  This relies on the  highly plausible conjecture that the complexity of approximating the permanent of Gaussian random matrices is \#$P$-hard. They showed that the probability distribution of submatrices of Haar random unitary matrices when $M \sim O(N^5)$ is close to a Gaussian distribution. Thus, it is possible to ``hide" a Gaussian random matrix inside a Haar random unitary matrix and this matrix determines the occupation number probability distribution of the Boson Sampler.  It is also believed that one can achieve this condition with a less stringent requirement, $M \sim O(N^2)$.

We thus seek to generate a Haar-random unitary transition matrix  in a 1D lattice.  In principle any transition matrix can be engineered by choosing an approximate hopping matrix $h_{l,l'}$, as can be achieved in the the coupling of arbitrary modes in a linear optical network, but for quantum random walkers, this would require highly nonlocal tunneling.  We show below that when restricted to only nearest-neighbor hopping in 1D, we can achieve an arbitrary transition matrix by allowing for a {\em time-dependent} lattice. By employing the tools of optimal control~\cite{Brif2010, Anderson2015} one can engineer any desired target unitary map on the system.  

The general optimal control paradigm is summarized as follows.   We consider a Hilbert space of finite dimension $d$ and a control Hamiltonian of the form
\begin{equation}
    H_c[\{\lambda^{(k)}\},t] = H_0 + \sum_k \lambda^{(k)}(t) H_k .
\end{equation}
The constant term is known as the ``drift" Hamiltonian and $\{\lambda^{(k)}(t)\}$ are the ``control  waveforms" that modulate other Hamiltonians $\{H_k\}$.   If the set $\{H_0, H_k\}$ is a generating set of the Lie algebra $\mathfrak{su}(d)$, then the system is said to be {\em controllable}~\cite{Jurdjevic1972, Brockett1973, Schirmer2002}.  That is for any target unitary $U_{tar} \in SU(d)$, there exists $\{\lambda_k(t)\}$ such that 
\begin{equation}
   U[\{\lambda^{(k)}\},T]= \mathcal{T}\left[e^{-i \int_0^T dt H_c[\{\lambda^{(k)}\},t] } \right] = U_{tar} 
\end{equation}
as long as $T \ge T_*$, where $1/T_*$ is known as the ``quantum speed limit"~\cite{Caneva2009}.  

One typically employs numerical methods to find $\{\lambda^{(k)}(t)\}$ by optimizing the fidelity $\mathcal{F}[\{\lambda^{(k)}\},T]=\vert Tr(U^\dag_{tar} U[\{\lambda^{(k)}\},T])\vert^2/d^2$ for some choice of $T$ sufficiently larger that $T_*$.  To do so, the waveforms $\{\lambda^{(k)}(t)\}$ are discretized by some representation and mapped to a vector of parameters $\vec{\lambda} $.  The minimal dimension of $\vec{\lambda}$ is $d^2-1$, the number of parameters to specify $U_{tar}\in SU(d)$. The choice of discretization depends on the trade-off between the ease of numerical optimization and the physical constraints on implementation of the waveform, e.g, bandwidth limitations.  For simplicity and proof-of principle, here we use a piecewise constant parameterization  $\vec{\lambda}=\{\lambda^{(k)}_i = \lambda^{(k)}(t_i)\; \vert \; t_i \leq t<t+i \delta t, i=1,\dots,K \}$, and $T=K \delta t$ for some choice of $\delta  t$ consistent with the physics of the control Hamiltonian. $K$ determines the dimension of $\vec{\lambda}$; $K_{min }=d^2-1$.   We employ a variant of the well known GRAPE algorithm to find optimal controls by gradient assent~\cite{GRAPE}.  These methods have been successfully employed in the Jessen group to design implement arbitrary high-fidelty SU(16) maps on the hyperfine magnetic sublevels of cesium atoms~\cite{Anderson2015}.

Here, we seek to employ optimal control to generate the complexity that is sufficient to achieve QCS.  This is feasible in a scalable way because we only need to design the {\em single-body} unitary map $u_{ll'}(T)$, an $M\times M$ matrix,  where $M$ is the number of modes included in the model, e.g., $M=50$ in the most ambitious case. That is, we seek to design a target transition matrix $\Lambda \in SU(M)$, picked according to the Haar measure, $U_{tar} = U_{Haar}$.  QCS is then achieved in the many-body system solely due to the quantum statistics of identical particles. This is in contrast to the similar efforts to observe QCS in quantum circuits with $N$ qubits with a series of quantum gates~\cite{google,Aaronson:16}.  There, one seeks to implement sufficiently random $2^N \times 2^N$ matrices, a task one cannot implement scalably via quantum control.

We consider two potential geometries that have been employed for studies of controlled transport: spinor lattices~\cite{Deutsch1998, Dur2002, Karski2009, Forster2009, Mischuck2010, Karski2011, Genske2013, Belmechri2013, Robens2016} and quantum gas microscopes~\cite{atomicgas, Sherson2010}. Each has different capabilities for optical control with experimentally accessible parameters.  We will analyze each in turn, proving controllability and numerically determining high-fidelty waveforms under realistic conditions.

\subsection{Spinor lattices}
\begin{figure}
\includegraphics[width=\columnwidth]{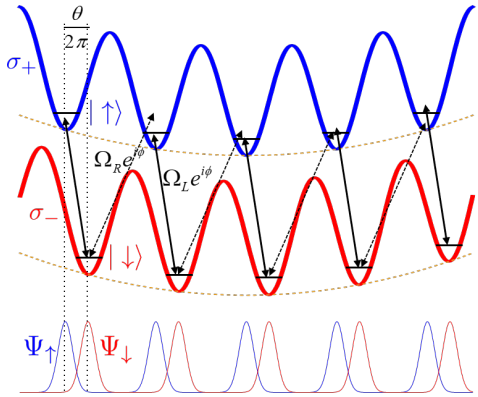}
\caption{ Coherent transport in a spinor optical lattice. A microwave field drives spin flips between two different hyperfine magnetic sublevels $\ket{\uparrow}$ and  $\ket{\downarrow}$, correlated with transport of the atomic wavepacket in the lowest Wannier states for optical lattices with $\sigma_\pm$ polarization.  The transition matrix element is determined by the wavepacket overlap (sketched below). The amplitude of hopping left(right) $\Omega_{L(R)}$ is controlled by the microwave amplitude and the time-dependent angle $\theta$, which determines the relative displacement of the two standing waves; the phase of the hoping matrix element is set by the  time-dependent microwave phase $\phi$.  The overall translational symmetry is broken by the addition of a quadratic chemical potential. } 
\label{fig:lattice}
\end{figure}

Without individual addressing of lattice sites, one way to achieve full control is to generalize the quantum random walk to spinor-optical lattices.  Specifically, consider two optical lattices with circular $\sigma_\pm$ polarization, whose nodes can be displaced by a parameter $\theta$, e.g., in a lin-angle-lin lattice~\cite{Deutsch1998,Dur2002, Forster2009}, or through direct synthesis of two independent standing waves~\cite{Robens2016}. Two chosen internal atomic states labeled $\ket{\uparrow}$ and  $\ket{\downarrow}$ can be trapped separately in each state-dependent lattice.  For concreteness we consider two magnetic sublevels in the electronic ground state separated in energy by the hyperfine splitting,  e.g., the stretched states, $\ket{\uparrow}=\ket{F=2,m_F=2}$, $\ket{\uparrow}=\ket{F=1,m_F=1}$, in  $^{87}$Rb.  Microwaves that drive spin flips between these internal states are then correlated with transport between the sublattices. We envision deep lattices along 1D so that tunneling of atoms within a given standing wave is highly suppressed and quantum walks occur only due to microwave-induced hopping in the tight-binding approximation (see Fig.~\ref{fig:lattice}).  We neglect here any interaction between the atoms, a good approximation if  the transverse confinement is weak, or by tuning the interaction to zero via a Feshbach resonance~\cite{Inouye1998}.

Our goal is to generate an arbitrary unitary transition matrix $\Lambda_{ll'}$.  In order to break the translational symmetry we add an additional quadratic light-shift potential. With this addition we show below  that the system is controllable when the relative phase of the standing waves and phase of the microwave are time-dependent control parameters.  We take as our control Hamiltonian

\begin{eqnarray}\label{ControlHamiltonia}
&H&_c[\theta,\phi ,t] = H_0 [\theta ,t]+ H_L[\theta,\phi ,t] + H_R[\theta, \phi ,t],  \\
&H&_0 [\theta,t] = V_0 \sum_l \left[\left(l-\Delta x_\downarrow \right)^2 a^\dag_{l,\downarrow} a_{l,\downarrow} \right. \nonumber \\ &&  \qquad \qquad \left. +  \left(l-\Delta x_\uparrow(\theta(t))\right)^2  a^\dag_{l,\uparrow} a_{l,\uparrow}\right], \nonumber\\
&H&_L[\theta, \phi, t] =  \sum_l \left[\frac{\Omega_L(\theta(t))}{2} e^{i\phi(t)} a^\dag_{l,\uparrow} a_{l,\downarrow}+h.c \right], \nonumber \\
&H&_R[\theta, \phi, t] =  \sum_l \left[ \frac{\Omega_R(\theta(t))}{2} e^{i\phi(t)} a^\dag_{l,\uparrow} a_{l+1,\downarrow}+h.c.\right]. \nonumber 
\end{eqnarray}

$H_0$ is a global quadratic potential added to the lattice potential, where $\Delta x_\downarrow = (M+1)/2$, and $\Delta x_\uparrow(\theta(t)) = (M+1)/2+\theta(t)/2\pi$.  $H_L$  and $H_R$ govern spin-dependent transport hopping to the left and right, driven by microwaves, with lattice labeling convention shown in Fig.~\ref{fig:lattice}. The control waveforms are denoted $\{ \theta(t), \phi(t)\}$. The hopping amplitudes are set by the global microwave Rabi frequency $\Omega_0$ and the spatial overlap of atomic wavepackets in the tight-binding approximation. For deep lattices, $\Omega_L(\theta(t)) =  \Omega_0 e^{-\left(\frac{\theta(t)}{2\eta}\right)^2}$,  $\Omega_R (\theta(t)) =  \Omega_0 e^{-\left(\frac{\theta(t)-\pi}{2\eta}\right)^2}$ where $\eta = k_L x_0$ is the Lamb-Dicke parameter, $x_0$ the width of the vibrational ground state Gaussian wavepacket, and $k_L$ is the lattice wavenumber~\cite{Forster2009}.  Additionally, the phase of the hopping amplitude is set by the global phase of the microwave, $\phi(t)$.  For this geometry we do not include any controls that address individual lattice sites.

 \begin{figure}
    \includegraphics[width=\columnwidth]{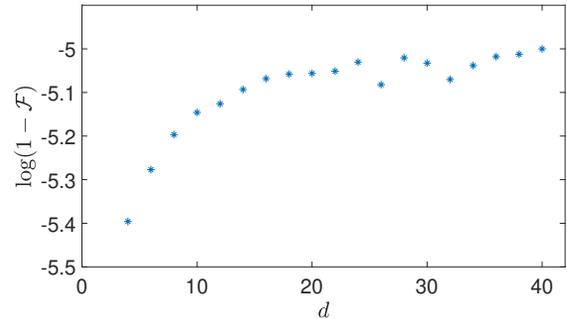}
    \caption{The infidelity between target unitary and unitary obtained from optimized control waveforms with respect to dimension for spinor lattices. Here the dimension of the Hilbert space is $d=2M$, where $M$ is the number of lattice sites. }
    \label{fig:infidelity}
\end{figure}

In Appendix A we prove this system is controllable, i.e., we can generate any $SU(M)\otimes U(2)$ matrix on a lattice with $M$ sites and two spin states.  With this in hand, we use numerical optimal control to find a time-dependent waveform that implements a desired unitary map as described above.  Here the target is a Haar random $SU(d)$ transition matrix, $d=2 M$. We parameterize the control waveforms as piecewise constant functions so that the Hamiltonian is a function of the control vector $\vec{\lambda} =\{ \theta (t_i),\phi (t_i) \vert i = 1,2,\dots , T/(\delta t)\}$, where $\delta t$ is the duration of each step, so that the number of steps is $K=T/\delta t$.  We take $K=d^2=4M^2$, slightly larger than the minimal parameterization to improve the stability of the `control landscape"~\cite{Moore2012}. The hopping frequency $J$ to set the scale of ``natural units" of our problem. For a lin-perp-lin lattice this is set by the Rabi frequency and wavepacket overlap, $\Omega_L(\theta=\frac{\pi}{2})=\Omega_R (\theta=\frac{\pi}{2})= \Omega_0$. We choose $\Omega_0 \delta t = 2\pi$, $T=K\delta t$, and  $\eta = 0.4$.

We maximize the fidelity function $\mathcal{F}(\vec{\theta},\vec{\phi}) = \left| \Tr \left(U_{Haar}^\dag U(\vec{\theta},\vec{\phi}) \right)\right|^2/d^2$. The optimization algorithm takes as its input the target Haar-random unitary map, a constant initial guess for the control waveform, and the form of gradient of fidelity based in the GRAPE algorithm.   
\begin{figure}
    \includegraphics[width=\columnwidth]{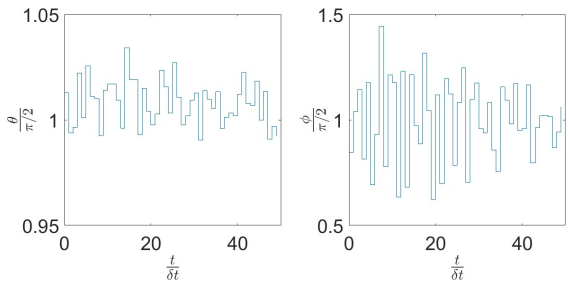}
    \caption{An example of a  proof-of-principle, piecewise constant control waveform that achieves a target Haar-random unitary for $d=2M=20$ (First 50 time steps). See text for the definition and choice of parameters.}
    \label{fig:cw}
\end{figure}
 Fig.~\ref{fig:infidelity} shows the average infidelity achieved for a collection of 20 Haar-random target unitary maps for different dimensions $M$.  For small lattices, the infidelity can be small as $10^{-5}$. Fig.~\ref{fig:cw} shows an example control waveform for $d=2M=20$.   Similar control waveforms have been implemented in the Jessen lab to achieve high-fidelity $SU(16)$ control on the hyperfine states of cesium~\cite{Anderson2015}.
 Note, in practice, laboratory implementations will be limited by finite slew rates and response bandwidth. In addition, high bandwidth can lead to ``leakage" outside the target Hilbert space.  Here the bandwith is set by $\Omega_0$.   As an example, for $\Omega_0/2\pi =100$ Hz, $\delta t = 10$ ms, this modulation time is sufficiently small to suppress heating for a typical trap oscillation frequency, of order 1 kHz.  The waveforms shown here demonstrate  a proof-of-principle construction.  In practice, one must design waveforms consistent with specific experimental constraints.
 
For ease of numerics and controllability, we have employed periodic boundary conditions, but this does not impose physical constraints. We initialize the system with $N$ atoms in $d$ lattice sites such that $d > N^2$. The optimization algorithm is restricted to a target unitary $M\times M$ $\Lambda$ matrix in a subspace of dimension $M =N^2$. If we trap the atoms in fixed $N$ lattice sites, the optimization will be a partial isometry with $N$ orthonormal columns in $d$-dimension. A general control waveform to achieve this target will require $dN$ time steps. If $d$ is chosen to be sufficiently larger than $M=N^2$, then the amplitudes will not reach the boundary in any intermediate step, and therefore will not have any boundary effects.

Numerical optimal control can never yield a fidelity exactly equal to one, and thus a perfect waveform to achieve a target Haar-random unitary is never possible. In addition, experimental errors will always occur in any implementation of a control waveform. However, as long as the final probability distribution is close to the ideal one in total variation distance, we can still guarantee its hardness~\cite{linoptics}.  Here, as long as the error that arises as the infidelity of generated unitary in our protocol is kept constant with respect to $M$, we can guarantee the hardness of output distribution~\cite{imperfection}. 

\subsection{Quantum gas microscopes}
A highly flexible system that can be used to achieve the requisite control is the quantum gas microscope. Atoms can be cooled and loaded with high fidelity into the ground state of a 2D optical lattices, and there is negligible mode mismatch in the tunnelling of atoms between sites. A potential candidate atom is $^{88}$Sr, which is a boson with nearly zero collision cross section at low temperature (scattering length $-2 a_0$) ~\cite{Mickelson2005}.  A 1D optical lattice with hard wall boundary conditions can be initialized with one atom in each of the desired sites.  The number of sites depends on the optical system of traps; a reasonable staring point is a 2D lattice of $30\times 30$ sites, or as many as $150-200$ sites in 1D~\cite{Kaufman_Privite}. The result is an ensemble of nearly identical noninteracting bosonic quantum random walkers.   Finally, with high fidelity one can also count the number of atoms in each site for $n=0,1,2$, the regime of interest for QCS, making this system ideal for Boson Sampling.  

Through the application of optical tweezers that are registered to the lattice, one can address individual sites~\cite{atomicgas,Norcia2018,Cooper2018}. These tweezers can be used to affect the depth of an individual lattice site, and/or the height of the barrier between sites.  These afford a tremendous degree of control for designing unitary map. We restrict our attention here to a 1D lattice with hard-wall boundary conditions. The control Hamiltonian takes the form 
\begin{eqnarray} \label{eq:AGM}
H(t)&=&H_x(t) + H_z(t) \\
H_x&=&\sum_{<l,l'>} h^x_{l,l'}(t) (a^\dag_l a_{l'}+ a^\dag_{l'} a_l) \\
H_z&=&\sum_{l} h^z_{l}(t) a^\dag_l a_l
\end{eqnarray}
Following from the proof for controllability in the spinor lattice, it is clear that this system is also controllable. This is shown in detail in Appendix B.

\begin{figure}
    \includegraphics[width=\columnwidth]{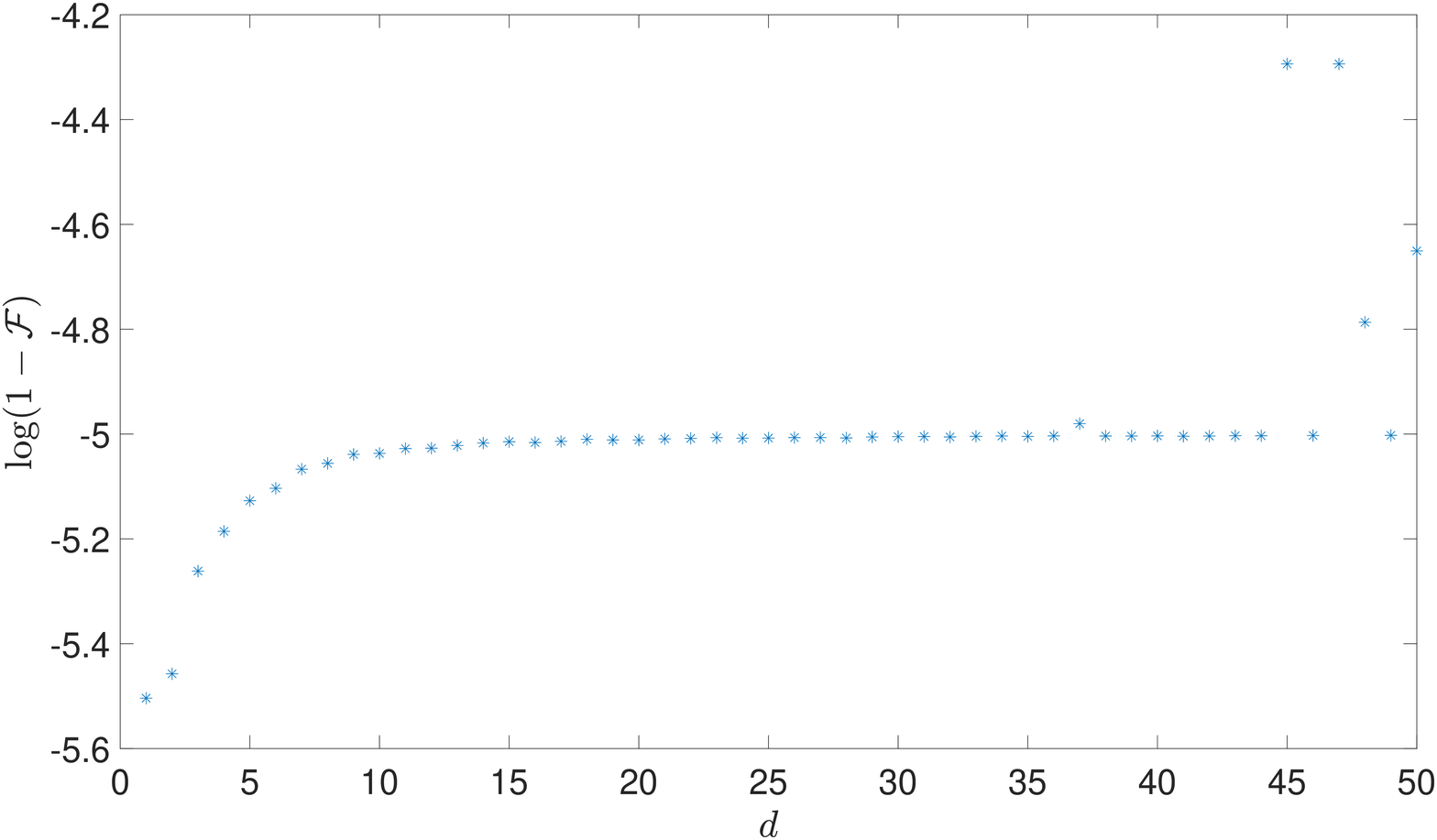}
    \caption{The infidelity between the target unitary map and unitary matrix obtained from optimized control waveforms with respect to the dimension for control with optical tweezers. The dimension $d=M$, where $M$ is the number of lattice sites.  See text for definitions of the parameters. }
    \label{fig:ot}
\end{figure}

\begin{figure}
    \includegraphics[width=\columnwidth]{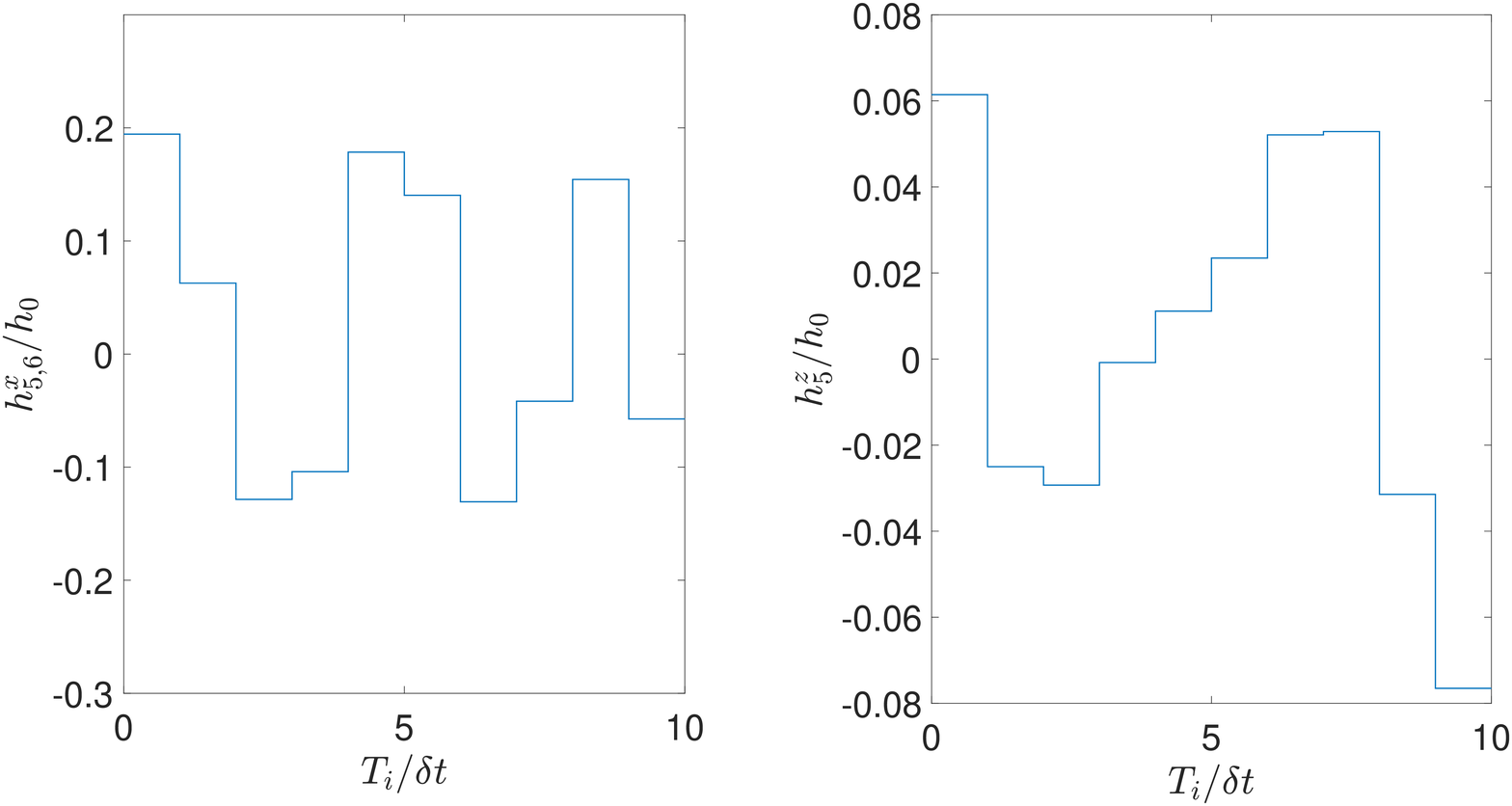}
    \caption{An example of a  proof-of-principle, piecewise constant control waveform that achieves a target Haar-random unitary in quantum gas microscope for $d=10$. Only one of each of $h^x_{l,l'}$ and  $h^z_{l}$ is shown here.  See text for the definition and choice of parameters.}
    \label{fig:wf}
\end{figure}

 In a similar fashion to what we showed for the spinor lattice, based on proof-of-principle piecewise constant waveforms one can achieve a high-fidelity implementation of the target Haar-random transition matrix.  We specify a separate vector of control amplitudes, individually addressed to each of $N$ lattice sites, $h^z_{l}(t_i)$, and each of $N-1$ tunneling couplings, $h^x_{l,l'}(t_i)$.  For each waveform we chose $N$ steps, each of duration $\delta t$, yielding a total of $2N^2 - N$ control parameters. As above, we numerically optimize the fidelity as a function of the control parameters using the modified GRAPE algorithm.  The fidelity averaged over 20 waveforms is shown in Fig.~\ref{fig:ot}.  Typical waveforms $h^z_{l}(t_i)$ and  $h^x_{l,l'}(t_i)$ are shown in Fig.~\ref{fig:wf} for $M=10$ lattice sites.  Here $h_0$ sets the characteristic hopping rate, and the scale for the controls, $h_0 \delta t = 2 \pi$.  As in the spinor lattice case the modulation must be designed to avoid atomic heating~\cite{Lacki2013}. 

\section{Summary and  outlook}
In this article, we have studied the potential for realizing QCS in the sampling complexity of an ensemble of noninteracting, identical bosonic quantum random walkers on a 1D lattice, restricted to near-neighbor hopping. For static lattices with uniform hopping, and for tunneling times at most logarithmic in the number of lattice sites, we specify an efficient classical Boson Sampling algorithm based on a calculation of the permanent of the transition matrix (to within  multiplicative approximation).  Surprisingly, for longer times, we know of no algorithm that can efficiently approximate the permanent, nor is  there any clear route to efficient Boson Sampling, even for the seemingly trivial problem of 1D Bloch bands in the tight-binding model.

While the 1D uniform case is not conclusive, we can obtain the complexity in Boson Sampling as conjectured by Aaronson and Arkhipov by implementing Haar random transition matrices.  One can achieve this with only nearest-neighbor hopping by allowing the lattice to be {\em time-dependent}.  Using the tools of quantum optimal control, one can implement a map that takes a state with a boson localized at one site to a target state that is a Haar-random superposition over all lattice sites. A map that simultaneously takes each localized boson to an orthogonal set of Haar-random states is a target Haar-random transition matrix.  Importantly, this is how  QCS is achieved by only controlling a {\em single-body map}, thanks to the exchange statistics of identical bosons. This is a sharp contrast to QCS based on random quantum circuits on (distinguishable) qubits.

We studied potential implementations of this protocol based on ultracold atomic bosons in optical potentials, a mature technology employed in studies of quantum simulations and other quantum information processing tasks.  In particular, we studied two trapping geometries: spinor optical lattices and quantum gas microscopes, each used in coherent transport experiments. We designed proof-of-principle control waveforms that can be used to implement high-fidelity Haar-random transition matrices with up to 40 lattice sites. Practical waveforms must take into account physical requirement such as the requirement to suppress heating of the atoms in the lattice (which would make the bosons distinguishable), and other constraints such as the bandwidth and slew-rate limitations of the controller, and finite coherence time in the system. 

Various tools for validation of the single particle unitary map can be borrowed from previous work on quantum optimal control on atomic hyperfine qudits.  Examples include randomized benchmarking~\cite{Anderson2015}, and efficient process tomography of near-unitary maps~\cite{Baldwin2014, sosa-martinez2017}. Given this validation, the operator norm between the target and measured single-body unitary map gives an upper bound on the total variation distance between the output probability distributions~\cite{Arkhipov2015}.  One can further verify the output probability distribution that include many-body interference using specially chosen unitary maps, such as the Fourier transform, where it is easy to predict that certain outputs are highly suppressed~\cite{Tichy2014,Dittel2018}.  The ultimate challenge is to verify highly complex distributions (speckle patterns in large dimensional Hilbert spaces). Current proposals involve choosing an intermediate problem size such that one can still use a classical supercomputer to generate samples, albeit with very large resources. One then performs statistical tests that compares the distributions, such as cross-entropy~\cite{Neill2018} or the so-called  HOG test~\cite{Neill2018}, which distinguishes the speckle pattern from a uniform distribution.

 Finally, while we have considered moderately sized systems, when looking ahead to scaling to even larger lattices, the tools of optimal control will be practically limited.  For example, a demonstration of QCS for 50 bosons in lattices of 2500 lattice sites would require optimal control design of $2500 \times 2500$ unitary transition matrix,  a task that would require powerful supercomputers. Indeed, while we have used quantum optimal control to reach a desired target Haar-random unitary, such fine-tuned control may not be required.  Since a random control waveform should yield a pseudorandom transition matrix after a suitably long mixing time (cf. ~\cite{Banchi2017}), we expect this time might be much shorter than that required to reach a particular target chosen from Haar random unitary matrices. This could lead to a more efficient and scalable approach to Boson Sampling.  It remains an open question as to how an  $\epsilon$-approximate $t$-design ~\cite{Oliveira2007,Marko2008,harrow2009,Dankert2009,Brown2010,brandao2016,Nakata2017,Harrow2018}, for sufficiently large $t$ and sufficiently small $\epsilon$, can accommodate approximate Gaussian random matrices necessary to achieve QCS in Boson Sampling.

We thank Tyler Keating, Christopher Jackson, Rolando Somma, Saleh Rahimi-Keshari, Alexey Gorshkov, and Daniel Burgath for helpful discussions of Boson Sampling and optimal control, and Adam Kaufman for detailed discussions relating to control in quantum gas microscopes. This work was supported in part by National Science Foundation grants PHY-1521016 and PHY-1521431.

\bibliographystyle{apsrev4-1}

%

\clearpage
\onecolumngrid
\appendix

\section{Proof of controllability for spinor lattices}
Our aim is to show that the Hamiltonian for the spinor lattice, Eq. (\ref{ControlHamiltonia}), is controllable in the one-body Hilbert space of $d=2M$ sites, $M$ lattice sites for each spin state. As discussed in the main text, one must prove that the set $\{H_0, H_R, H_L\}$ form the generators of the Lie algebra $\mathfrak{su}(d)$.  Since we are considering a spinor lattice with two spin states, $\mathfrak{su}(d)=\mathfrak{su}(M)\otimes\mathfrak{u}(2)$.  A basis for this algebra is $\{G_x,G_y,G_z\} \otimes \{ \mathbb{1},\sigma_x,\sigma_y,\sigma_z \}$ where $\{G_x,G_y,G_z\}$ are generalized Gell-Mann Matrices, which are the basis for any (traceless) Hermitian operator (here in $M$ dimensions.  The generalized Gell-Mann Matrices are as follows:
 \begin{itemize}
     \item Symmetric: $G_x^{jk}= |j\rangle\langle k| + |k\rangle\langle j|$
     \item Anti-symmetric: $G_y^{jk}= -i|j\rangle\langle k| + i |k\rangle\langle j|$
     \item Diagonal: $G_z^{l}= \sum\limits_{j=1}^{l} |j\rangle\langle j| - l |l+1\rangle\langle l+1|$
 \end{itemize}
 Let us also define $G_{\phi}^{j,k}= e^{i\phi}|j\rangle\langle k| + e^{-i\phi} |k\rangle\langle j|. $
 Our strategy will be the following. The list of Hamiltonians that we have control over is in Eq. (~\ref{ControlHamiltonia}). We are allowed to take multiple nested commutators and any linear combinations of terms in our list. Every time we obtain a new term we add it to our list. We will similarly expand the list until it has all the Gell-Mann matrices. First we will obtain the nearest neighbor terms of the form $G_{\phi}^{l,l+1}\otimes\{\mathbb{1},\sigma_z\}$ and add it to our list. Then we will show that we can obtain all Gell-Mann matrices from the nearest neighbor terms. 
 We shall now write down commutators of different terms in the Hamiltonian, which are relevant in obtaining terms of our choice. Henceforth, we will omit the overall constant factors, such as $i$, that appear in commutators and use the symbol $\Rightarrow$ denote the new terms that arise in the generating set, from which we can take linear combinations and further nested commutators.  Consider, thus,
 \begin{eqnarray}
[H_L,H_R] &\Rightarrow& \sum\limits_{l} G_{\phi}^{l+1,l} \otimes \sigma_z \nonumber \\ 
    \left[H_L,\left[H_R,H_0 \right]\right] &\Rightarrow& \sum\limits_{l=1}^{M-1} (1-\frac{\theta}{\pi})(2l+\frac{\theta}{\pi}-M) G_{\phi}^{l.l+1} \otimes \sigma_z  + \frac{\theta}{\pi} (M+\frac{\theta}{\pi}-1) G_{\phi}^{M,1} \otimes \sigma_z  \label{eq:GM} \\
\left[H_L(\phi=0),H_L(\phi=\frac{\pi}{2})\right]&\Rightarrow&\mathbb{1}\otimes\sigma_z \nonumber
\end{eqnarray}
Substituting $\theta=\pi$ in the Eq. (~\ref{eq:GM}), we obtain
\begin{eqnarray*}
\left[H_L,\left[H_R,H_0 \right]\right]  \Rightarrow G_{\phi}^{M,1} \otimes \sigma_z .
 \end{eqnarray*}
 By taking the following linear combination, and commutator
 \begin{eqnarray*}
H_0(\theta=\pi)-H_0(\theta=0) &=& \Big( (\frac{1}{4} - \Delta_{x_{\downarrow}}) \mathbb{1} + \sum\limits_l l |l\rangle\langle l| \Big) \otimes|\downarrow\rangle\langle\downarrow |, \nonumber
\\ &\equiv& \hat{Z}\otimes |\downarrow\rangle\langle\downarrow | 
\end{eqnarray*}
we can obtain the following term
\begin{equation*}
    \left[ \left[ \hat{Z}\otimes |\downarrow\rangle\langle\downarrow |,\mathbb{1} \otimes \sigma_x\right],\mathbb{1} \otimes \sigma_x\right] \Rightarrow\hat{Z}\otimes \sigma_z . \nonumber
\end{equation*}
By taking an additional commutator, we isolate a desired class of generators,
\begin{equation*}
\left[ \hat{Z}\otimes \sigma_z, G_{\phi-\pi/2}^{M,1} \otimes \sigma_z\right] \Rightarrow G_{\phi}^{M,1} \otimes\mathbb{1}.
\end{equation*}

At this point we have obtained all the set of terms of the form $G_{\phi}^{M,1}\otimes\{\mathbb{1},\sigma_z\}$. We can see that $G_{0}^{M,1}=G_{x}^{M,1}$ and $G_{\pi/2}^{M,1} =G_{y}^{M,1}$, which are two of the Gell-mann matrices. Using these can can generate terms of the form $G_{\phi}^{l,l+1}\otimes\{\mathbb{1},\sigma_z\}$.  To do so, we note 
\begin{equation*}
     \left[G_{x}^{M,1} \otimes \sigma_z,G_{y}^{M,1}\otimes \sigma_z \right] \Rightarrow |1\rangle\langle 1| - |M\rangle\langle M| 
\end{equation*}
\begin{equation*}
     H_0(\theta=0) = \sum\limits_{l} (l-\Delta_{x_{\downarrow}})^2 |l\rangle \langle l| \otimes \mathbb{1}
\end{equation*}
 \begin{equation*}
  \left[ \sum\limits_{l} G_{\phi}^{l+1,l} \otimes \sigma_z, |1\rangle\langle 1| - |M\rangle\langle M|\right] \Rightarrow \hat{B}(\phi) \otimes \sigma_z,
  \end{equation*}
 where 
  \begin{equation*}
  \hat{B}(\phi) = G_{\phi}^{1,2} -2 G_{\phi}^{M,1} + G_{\phi}^{M-1,M}.
  \end{equation*}
 Since we already have the term $G_{\phi}^{M,1}\otimes \sigma_z$ in our set, we obtain the term $\hat{C}=\left(G_{\phi}^{1,2} + G_{\phi}^{M-1,M}\right)\otimes \sigma_z$ from the above commutator. Now,
 \begin{equation*} \label{eq:C}
  \left[ \hat{C}, H_0(\theta=0) \right] \Rightarrow \left( (2-M)  G_{\phi}^{1,2}+ M G_{\phi}^{M-1,M} \right) \otimes \sigma_z.
 \end{equation*}
The operators in this equation and $\hat{C}$ are linearly independent in $G_{\phi}^{1,2}$, and $G_{\phi}^{M-1,M}$ . Thus, we can obtain those terms separately through an appropriate linear combination. In a similar way, whenever we have access to generators that are sum of new terms and the terms of which we already have access, then we can use linear combination to extract the new terms.  We use the symbol $\Rightarrow$ to include both commutators and linear combinations.  Thus,
  \begin{eqnarray*}
  \Bigg[ G_{x}^{1,2}\otimes \sigma_z, G_{y}^{1,2}\otimes \sigma_z\Bigg] &\Rightarrow&\hat{A}^z_{1}= |2\rangle\langle 2| - |1\rangle\langle 1| \\
  \Bigg[ \sum\limits_{l} G_{\phi}^{l+1,l} \otimes \sigma_z, \hat{A}^z_{1}\Bigg] &\Rightarrow&\left(G_{\phi}^{2,3} + 2 G_{\phi}^{1,2} + G_{\phi}^{M,1}\right)\otimes \sigma_z \nonumber\\  &\Rightarrow& G_{\phi}^{2,3}\otimes \sigma_z  \\
  \Bigg[ G_{x}^{2,3}\otimes \sigma_z ,G_{y}^{2,3}\otimes \sigma_z \Bigg] &\Rightarrow& \hat{A}^z_{2}=|3\rangle\langle 3| - |2\rangle\langle 2|\\
 \Bigg[ \sum\limits_{l} G_{\phi}^{l+1,l} \otimes \sigma_z, \hat{A}^z_{2}\Bigg] &\Rightarrow& \left(G_{\phi}^{3,4} + 2 G_{\phi}^{2,3} + G_{\phi}^{1,2} \right)\otimes \sigma_z \nonumber\\ &\Rightarrow& G_{\phi}^{3,4}\otimes \sigma_z \\
  &&\vdots \nonumber
  \end{eqnarray*}
 If we repeat the above steps, we obtain all terms of the form $G_{\phi}^{l,l+1}\otimes \sigma_z$. Repeating again, starting from $G_{\phi}^{M,1}\otimes \mathbb{1}$, we obtain all terms of the form $G_{\phi}^{l,l+1}\otimes \mathbb{1}$.

 Next we will show that terms of the form $G_{\phi}^{l,l+1}\otimes\{ \mathbb{1},\sigma_z\} $ are sufficient to generate all of the Gell-Mann matrices. Notice that $ \left[G_{\phi}^{l,l+1}\otimes \sigma_z,  \mathbb{1}\otimes \sigma_{\phi'-\pi/2}\right] \Rightarrow G_{\phi}^{l,l+1}\otimes \sigma_{\phi'}$. Without loss of generality we can assume that $j<k$. We can prove this using mathematical induction. We have following terms,
 \begin{eqnarray} \label{eq:Gell}
 G_{\phi}^{j,j+1}\otimes \mathbb{1}&&  \\
 G_{\phi}^{j,j+1}\otimes \sigma_z && \nonumber\\
 \left[G_{\phi}^{j,j+1}\otimes \sigma_z, \mathbb{1} \otimes \sigma_{\phi'-\frac{\pi}{2}}\right] &\Rightarrow& G_{\phi}^{j,j+1}\otimes \sigma_{\phi'} \nonumber
 \end{eqnarray}
 Now suppose we have access to $G_{\phi}^{j,k}\otimes \{\mathbb{1},\sigma_z,\sigma_{\phi'}\}$. We want to show that we can generate $G_{\phi}^{j,k+1}\otimes \{\mathbb{1},\sigma_z,\sigma_{\phi'}\}$.
 \begin{eqnarray*}
 G_{\phi}^{j,k+1} \otimes \mathbb{1} &=& \left(e^{-i\phi}|j \rangle\langle k+1| + e^{i\phi} |k+1 \rangle\langle j| \right) \otimes \mathbb{1} \\
 &\Rightarrow&  \left[\left(e^{-i(\phi+\frac{\pi}{2})}|j \rangle\langle k| + e^{i(\phi+\frac{\pi}{2})}|k \rangle\langle j| \right) \otimes \mathbb{1},\left(|k \rangle\langle k+1| + |k+1 \rangle\langle k| \right) \otimes \mathbb{1} \right]\\
&\Rightarrow& \left[ G_{\phi+\frac{\pi}{2}}^{j,k}\otimes \mathbb{1},\hat{A}_{k}(0)\otimes \mathbb{1}\right]\\ 
\end{eqnarray*}
and {similarly}
 \begin{eqnarray*}
 G_{\phi}^{j,k+1} \otimes\{ \sigma_z,\sigma_{\phi'}\} &=& \left(e^{-i\phi}|j \rangle\langle k+1| + e^{i\phi} |k+1 \rangle\langle j| \right) \otimes \{ \sigma_z,\sigma_{\phi'}\} \\
 &\Rightarrow&  \left[\left(e^{-i(\phi+\frac{\pi}{2})}|j \rangle\langle k| + e^{i(\phi+\frac{\pi}{2})}|k \rangle\langle j| \right) \otimes \mathbb{1},\left(|k \rangle\langle k+1| + |k+1 \rangle\langle k| \right) \otimes \{ \sigma_z,\sigma_{\phi'}\} \right] \\
 &\Rightarrow&  \left[ G_{\phi+\frac{\pi}{2}}^{j,k}\otimes \mathbb{1},\hat{A}_{k}(0)\otimes \{ \sigma_z,\sigma_{\phi'}\} \right]
 \end{eqnarray*}
 
\begin{eqnarray*}
  \left[ G_x^{jk} \otimes \mathbb{1} ,G_y^{jk} \otimes  \{\mathbb{1},\sigma_z, \sigma_{\phi}\} \right]  &\Rightarrow& \left(|j\rangle\langle j| - |k\rangle\langle k|\right) \otimes \{\mathbb{1},\sigma_z, \sigma_{\phi}\} \\
  G_z^{l} \otimes \{\mathbb{1},\sigma_z,\sigma_{\phi}\}& = &  \sum\limits_{j=1}^{l} \left( |j\rangle\langle j| - |l+1\rangle\langle l+1| \right)\otimes \{\mathbb{1},\sigma_z, \sigma_{\phi}\}
 \end{eqnarray*}
 We have thus proven that this is sufficient to generate all of the Gell-Mann matrices of a system of spinor optical lattices.

\section{Proof of controllability for atomic gas microscopy using optical tweezers}

Here our aim is show that the Hamiltonian for quantum gas microscope in Eq.~(\ref{eq:AGM}) is controllable. The dimension is $d=M$, where $M$ is the number of lattice sites. We try to generate all Gell-Mann matrices (as defined in previous section). The terms that we have access to are 

\begin{eqnarray*}
G_x^{l,l+1}&=& |l\rangle\langle l+1| + |l+1\rangle\langle l| \\
\tilde{G}_z^{l}&=& |l\rangle\langle l|
\end{eqnarray*}
We can take following commutators and linear combination to obtain some new terms.
\begin{eqnarray*}
\left[G_x^{l,l+1},\tilde{G}_z^{l}\right] &=&|l\rangle\langle l+1| - |l+1\rangle\langle l| \\ &\Rightarrow&  G_y^{l,l+1} \\
\cos(\phi) G_x^{l,l+1} + \sin(\phi) G_y^{l,l+1} &=& G_{\phi}^{l,l+1}
\end{eqnarray*}
Now we have access to all terms of the form $G_{\phi}^{l,l+1}$. Now following all steps from Eq.(~\ref{eq:Gell}), we can easily show that it generates all Gell-Mann matrices.
\end{document}